\documentclass[journal]{IEEEtran}
\usepackage{amsmath}
\usepackage{graphicx}
\usepackage{wrapfig}
\begin{document}
\title{Performance Analysis of a Cognitive Radio Network with an Energy Harvesting Secondary Transmitter under Nakagami-{\textit{m}} Fading}
\author{Binh Van Nguyen, Hyoyoung Jung, Dongsoo Har, and Kiseon Kim
\thanks{Binh Van Nguyen, Hyoyoung Jung, and Kiseon Kim are with the School of Electrical Engineering and Computer Science, Gwangju Institute of Science and Technology, Republic of Korea.
(E-mail : \{binhnguyen, rain, kskim\}@gist.ac.kr).}
\thanks{Dongsoo Har is with the Cho Chun Shik Graduate School for Green Transportation, Korea Advanced Institute of Science and Technology, Republic of Korea.
(E-mail: dshar@kaist.ac.kr).}
\thanks{The authors gratefully acknowledge the support from Electronic Warfare Research Center at Gwangju Institute of Science and Technology (GIST), originally funded by Defense Acquisition Program Administration (DAPA) and Agency for Defense Development (ADD).}
}
\maketitle
\begin{abstract}
In this paper, we consider an overlay cognitive radio network, in which a secondary transmitter (ST) is willing to relay the information of a primary transmitter toward a primary receiver. In return, ST can access the licensed band to send its own information superimposed with the primary signal to a secondary receiver. The power-limited ST uses a power splitting protocol to harvest energy from its received signal to increase its transmit power. We analyze the performance of the primary and the secondary systems under independent Nakagami-{\textit{m}} fading by deriving their corresponding outage probabilities in integral-based expressions. In addition, by considering the high signal-to-noise ratio, we obtain very tight closed-form approximations of the outage probabilities. Thereafter, by further analyzing the approximations, we reveal novel insights on the diversity orders and coding gains of the two systems. Our analytical results are validated through extensive Monte-Carlo simulations.
\end{abstract}


\IEEEpeerreviewmaketitle

\section{Introduction}
% Energy Harvesting & CRNs
\IEEEPARstart{E}{nergy} harvesting (EH) has been proven as a promising solution to prolong the lifetime of energy-constraint wireless communication systems. In addition, beside conventional EH methods, i.e. harvesting energy from wind, solar, and vibration, harvesting energy from radio frequency signals (RF-based EH) is emerging as a favorable alternative \cite{Fafoutis'15}. The importance of RF-based EH lies in the fact that RF signals carry both information and energy at the same time, and thus, RF-based EH allows limited-power nodes to scavenge energy and process information simultaneously \cite{Wang'16}. On the other hand, cognitive radio is a powerful mean to solve the problem of wireless spectrum scarcity and licensed spectrum under-utilization \cite{Liang'11}. In addition, with the rapid growth of Internet-of-Things (IoT), applications based on cognitive ratio networks (CRNs) are expected to be widely spread in the near future \cite{Khan'16}.

% ENERGY HARVESTING IN CRNs
Considering RF-based EH in CRNs has recently got a lot of attraction from research community, i.e. \cite{Hoang'14}-\cite{Jain'15} and references therein. More specifically, in \cite{Hoang'14}, a channel access policy is proposed to maximize a secondary throughput of an interweave CRN. In addition, in \cite{Liu'16}, outage probability (OP) and throughput of an underlay CRN is investigated. Moreover, in \cite{Zheng'14}, information and energy cooperations among primary users (PUs) and secondary users (SUs) in an overlay CRN is studied from an optimization perspective. Furthermore, the authors of \cite{Jain'15} analyze the OP of an overlay CRN under the Nakagami-{\it m} channels. It is noted that the overlay paradigm is typically preferred for scenarios in which a primary system itself cannot achieve a satisfactory quality of service due to severe fading. In such cases, SUs can help PUs to obtain a better primary performance, and in return, SUs can harvest energy and access the primary spectrum to exchange their own information. This is a win-win cooperation which is expected to be widely spread in future RF-based EH networks.

In the last paper above, the OPs of primary and secondary systems are provided in complex expressions, hence, it is not easy to understand the systems behaviors and the effect of key parameters on the systems' performances. Motivated by this fact, in this paper, we first extend the system configuration of \cite{Jain'15} by including the direct primary transmitter (PT) - primary receiver (PR) channel and then derive the OPs of the two systems in integral-based expressions. In addition, and importantly, we carry out tractable closed-form approximations of the primary and secondary OPs. From the derived approximations, we are able to reveal novel insights about the primary and secondary diversity orders (DOs) and coding gains (CGs). In addition, our results provide a guideline how to select the value of the power-sharing coefficient used for superimposing primary and secondary signals depending upon which system performance is preferred.

The remainder of this paper is organized as follows. Section II introduces the network and the channel models. Exact primary and secondary OPs are derived in Section III. Approximations of the OPs, DOs, and CGs of the primary and the secondary systems are presented in Section IV. Simulation results are provided in Section V, followed by our conclusion in Section VI.

\section{Network and Channel Models}
\subsection{Network Model}
\begin{figure}[t]
    \centering
    \includegraphics[width = 7.5cm]{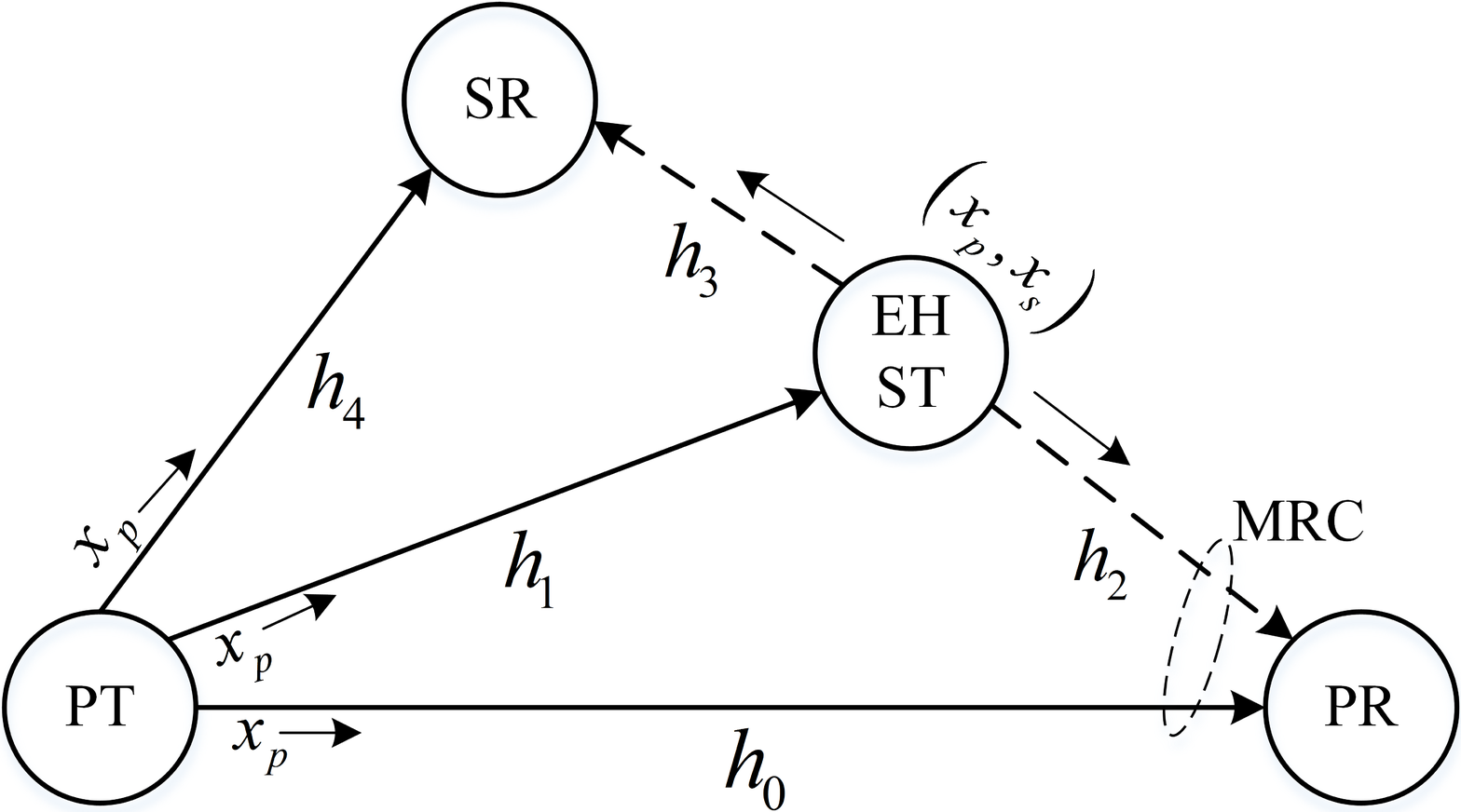}
    \caption{An overlay cognitive radio network with a RF-based EH secondary transmitter.}
\end{figure}
\begin{figure*}[!t]
    \centering
    \includegraphics[width = 11cm]{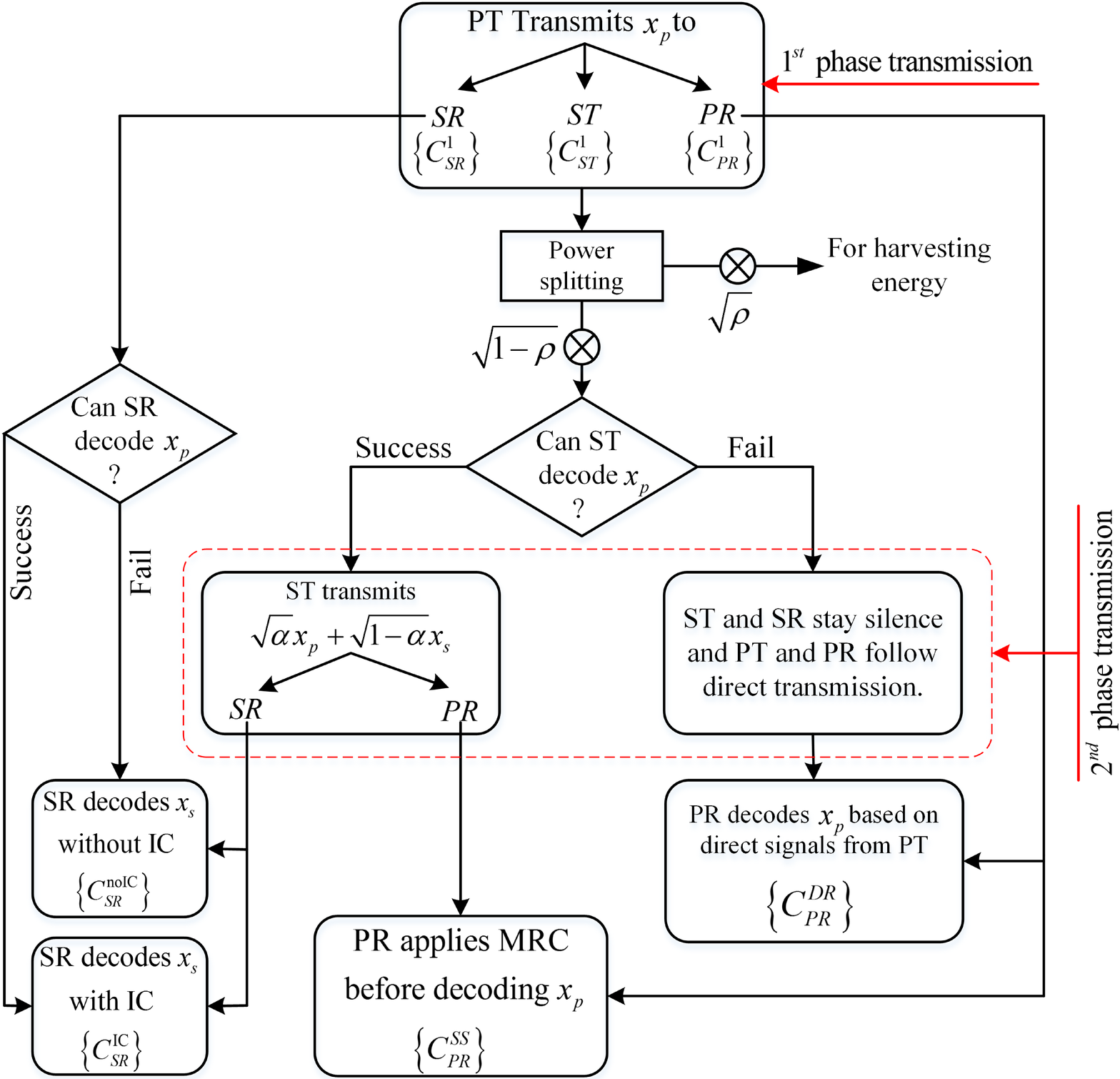}
    \caption{Network transmission.}
\vspace*{2pt}
\hrulefill
\end{figure*}
The considered network consists of primary and secondary systems, as illustrated in Fig. 1, shown as the top of the next page. The former system comprises a PT-PR pair, while the latter one consists of a secondary transmitter-secondary receiver (ST-SR) pair. Each node has a single antenna and operate in the half-duplex mode. In addition, ST is assumed to be a self-sustaining node with the RF-based EH capability. Moreover, ST deploys the power-splitting EH protocol to harvest energy from its received signals.

Each transmission takes place in two phases, each of which has a duration $T$ in time unit. In the first phase, PT broadcasts primary signal $x_p$ with $E\left[ {{{\left| {{x_p}} \right|}^2}} \right] = 1$ to PR and ST, which also be overheard by SR. $E\left[ \cdot \right]$ here denotes the expectation operator. The received signal at node $i$ is expressed as follows
\begin{align}
y_i^1 = \sqrt P {h_j}{x_p} + {n_i},
\end{align}
where $i = \left\{ {PR,ST,SR} \right\}$, $j = \left\{ {0,1,4} \right\}$, $P$ is the transmit power of PT, $h_j$ denotes the channel coefficients, and $n_i$ is the zero-mean and $N_0$ variance AWGN at node $i$. In (1), the superscript "1" denotes the first phase. Thereafter, ST divides $y_{ST}^1$ into two streams by a power splitting coefficient $0 \le \rho \le 1$. Particularly, $\sqrt{\rho} y_{ST}^1$ is for harvesting energy and $\sqrt{1 - \rho} y_{ST}^1$ is for decoding $x_p$. Consequently, the harvested energy at ST is given by
\begin{align}
E_{ST} = \rho \eta P T g_1,
\end{align}
where ${g_j} = {\left| {{h_j}} \right|^2}$ and $0 < \eta \leq 1$ is the energy conversion efficiency. In addition, taking additive noises generated by RF-to-baseband down conversion into account, we have
\begin{align}
& y_{PR}^1 = \sqrt P {h_0}{x_p} + {n_{PR}} + n_{PR}^c, \\
& y_{SR}^1 = \sqrt P {h_4}{x_p} + {n_{SR}} + n_{SR}^c, \\
& \left( 1-\rho \right) y_{ST}^{1} = \sqrt {\left( {1 - \rho } \right)P} {h_1}{x_p} + \sqrt {1 - \rho } {n_{SR}} + n_{ST}^c,
\end{align}
where $n_i^c$ denotes the conversion noise which is assumed to be zero-mean AWGN with variance $\mu N_0$.

If ST fails to decode $x_p$, it will stay silence in the second phase and PUs will follow the direct PT-PR transmission. Otherwise, ST will superimpose $x_p$ with its own signal $x_s$ and transmit $\sqrt \alpha  {x_p} + \sqrt {1 - \alpha } {x_s}$ to PR and SR in the second phase. $0 \le \alpha \le 1$ here is the power sharing coefficient. The received signal at PR and SR are written as follows
\begin{align}
y_{PR}^2 = \sqrt {\alpha {P_{ST}}} {h_2}{x_p} + \sqrt {\left( {1 - \alpha } \right){P_{ST}}} {h_2}{x_s} + \tilde{n}_{PR}, \\
y_{SR}^2 = \sqrt {\left( {1 - \alpha } \right){P_{ST}}} {h_3}{x_s} + \sqrt {\alpha {P_{ST}}} {h_3}{x_p} + \tilde{n}_{SR},
\end{align}
where $\tilde{n}_{PR}=n_{PR} + n_{PR}^c$, $\tilde{n}_{SR}=n_{SR} + n_{SR}^c$, and $P_{ST} = E_{ST}/T = P \rho \eta g_1$ is the transmit power of ST.  Here, we assume that the harvested energy is always larger than the circuit activation energy. In addition, we assume that ST dispenses all of its harvested energy for information transmission during the second phase \cite{Liu'16}-\cite{Jain'15}.

After receiving $y_{PR}^2$, PR combines $y_{PR}^1$ and $y_{PR}^2$ for decoding $x_p$. In addition, since SR overhears the primary signal in the first phase, it may exploit the overheard signal to cancel the interference term, $\sqrt {\alpha {P_{ST}}} {h_3}{x_p}$, contained in $y_{SR}^2$. As a result, SR can decode $x_s$ either with or without interference cancellation (IC). Finally, from (3)-(7) the capacities obtained at SR, ST, and PR in the first phase ($C_{SR}^1$, $C_{ST}^1$, $C_{PR}^1$), at PR with direct and cooperative transmissions ($C_{PR}^{PT \rightarrow PR}$, $C_{PR}^{PT \rightarrow ST \rightarrow PR}$), and at SR with and without IC in the second phase ($C_{SR}^{\text{with} \; IC}$, $C_{SR}^{\text{without} \; IC}$) are derived as follows
\begin{align}
\label{C-SR-1}
& C_{SR}^1 = \frac{1}{2}{\log _2}\left( {1 + \frac{{\bar \gamma {g_4}}}{{1 + \mu }}} \right), \\
\label{C-ST-1}
& C_{ST}^1 = \frac{1}{2}{\log _2}\left( {1 + \frac{{\left( {1 - \rho } \right)\bar \gamma {g_1}}}{{1 - \rho  + \mu }}} \right), \\
\label{C-PR-1}
& C_{PR}^1 = \frac{1}{2}{\log _2}\left( {1 + \gamma _{PR}^1} \right), \\
\label{C-PR-CSS}
& C_{PR}^{PT \rightarrow ST \rightarrow PR} \nonumber \\
&= \frac{1}{2}{\log _2}\left( {1 + \gamma _{PR}^1 + \frac{{\alpha \bar \gamma \rho \eta g_1 g_2}}{{\left( {1 - \alpha } \right)\bar \gamma \rho \eta g_1 g_2 + 1 + \mu }}} \right), \\
\label{C-SR-noIC}
& C_{SR}^{\text{without} \;IC} = \frac{1}{2}{\log _2}\left( {1 + \frac{{\left( {1 - \alpha } \right)\bar \gamma \rho \eta g_1 g_3}}{{\alpha \bar \gamma \rho \eta g_1 g_3 + 1 + \mu }}} \right), \\
\label{C-SR-IC}
& C_{SR}^{\text{with} \;IC} = \frac{1}{2}{\log _2}\left( {1 + \frac{{\left( {1 - \alpha } \right)\bar \gamma \rho \eta g_1 g_3}}{{1 + \mu }}} \right), \\
\label{C-PR-DR}
& C_{PR}^{PT \rightarrow PR} = {\log _2}\left( {1 + \gamma _{PR}^1} \right),
\end{align}
where $\bar \gamma  = P/{N_0}$ and $\gamma _{PR}^1 = \frac{{\bar \gamma {g_0}}}{{1 + \mu }}$. A detailed graphically illustration of the next work's transmission is provided in the Fig. 2, shown at the top of the next page.

\subsection{Channel Model}
The channel coefficients $h_j$ are assumed to be independent Nakagami-{\textit{m}} random variables (RVs). The probability density function (PDF) of a Nakagami-{\textit{m}} RV $X$ is given by
\begin{align}
{f_X}\left( x \right) = \frac{{2{m^m}{x^{2m - 1}}}}{{\Gamma \left( m \right){\beta ^m}}}{e^{\frac{{ - m}}{\beta }{x^2}}},
\end{align}
where $\beta  = E\left[ {{X^2}} \right]$ denotes the variance of $X$, $\Gamma \left(  \cdot  \right)$ denotes the Gamma function, and $m$ is the Nakagami fading figure that controls the depth of the fading envelope. The above distribution models deep fading when $0.5 \le m  < 1$, Rayleigh fading with $m = 1$. When $m > 1$, it gives the fading that is less severe than Rayleigh. Furthermore, when $m = \infty$, it simplifies to AWGN \cite{Guffin'03}. The fading figures and variances of $h_0$, $h_1$, $h_2$, $h_3$, and $h_4$ are respectively denoted as $\left\{ {m_0,\beta_0} \right\}$, $\left\{ {m_1,\beta_1} \right\}$, $\left\{ {m_2,\beta_2} \right\}$, $\left\{ {m_3,\beta_3} \right\}$, and $\left\{ {m_4,\beta_4} \right\}$.

From the above channel model, we can readily verify that $g_j = {\left| {{h_j}} \right|^2}$ for $j = \left\{ {1,2,3,4} \right\}$ follows the Gamma distribution with PDF and cumulative density function (CDF) are given by
\begin{align}
\label{eq:f_Ga}
& {f_{{g_j}}}\left( x \right) = \frac{{{x^{{m_j} - 1}}{e^{ - \frac{x}{{{\Omega _j}}}}}}}{{{\Omega _j}^{{m_j}}\Gamma \left( {{m_j}} \right)}}, \\
& {F_{{g_j}}}\left( x \right) = \frac{{\gamma \left( {{m_j},x/{\Omega _j}} \right)}}{{\Gamma \left( {{m_j}} \right)}},
\end{align}
where ${\Omega _j} = {\beta _j}/{m_j}$ and $\gamma \left( { \cdot , \cdot } \right)$ denotes the lower incomplete Gamma function which is defined as $\gamma \left( {a,x} \right) = \int\limits_0^x {{t^{a - 1}}{e^{ - t}}dt}$.

\section{Exact Outage Probability Analysis}
In this section, we derive the exact expressions of the OPs of the primary and secondary systems. The derived results are useful for us to quickly evaluate the systems performance without doing complex computer simulations.

\subsection{Outage Probability of the Primary System}
We first start with the OP of the primary system which occurs when 1) $C_{ST}^1 < R_0$ and $C_{PR}^{PT \rightarrow PR} < R_0$ or 2) $C_{ST}^1 \geq R_0$ and $C_{PR}^{PT \rightarrow ST \rightarrow PR} < R_0$, where $R_0$ is a pre-defined target rate in bits/s/Hz. In other words, the OP of the primary system can be formulated as follows
\begin{align} \label{OP-1}
O{P_{p}} & = \underbrace {\Pr \left[ {{C_{ST}^1} < {R_0},{C_{PR}^{PT \rightarrow PR}} < {R_0}} \right]}_{I_1} \nonumber \\
& \;\;\;\; + \underbrace {\Pr \left[ {{C_{ST}^1} \ge {R_0},{C_{PR}^{PT \rightarrow ST \rightarrow PR}} < {R_0}} \right]}_{I_2},
\end{align}
where $I_1$ and $I_2$ can be further rewritten as
\begin{align}\label{Pr1}
I_1 & = \Pr \left[ {{g_1} < \frac{{T_1\left( {1 - \rho  + \mu } \right)}}{{\left( {1 - \rho } \right)\bar \gamma }},{g_0} < \frac{{T_0\left( {1 + \mu } \right)}}{{\bar \gamma }}} \right], \nonumber \\
& = \frac{{\gamma \left( {{m_1},{\theta _1}/{\Omega _1}} \right)\gamma \left( {{m_0},{\theta _0}/{\Omega _0}} \right)}}{{\Gamma \left( {{m_1}} \right)\Gamma \left( {{m_0}} \right)}}.
\end{align}
and
\begin{align}
{I_2} = \Pr \left[ {{g_1} \ge \frac{{{T_1}\left( {1 - \rho  + \mu } \right)}}{{\left( {1 - \rho } \right)\bar \gamma }},\gamma _{PR}^1 + \gamma _{PR}^2 < {T_1}} \right],
\end{align}
with $\gamma _{PR}^2 = \frac{{\alpha \bar \gamma \eta \rho {g_1}{g_2}}}{{\left( {1 - \alpha } \right)\bar \gamma \eta \rho {g_1}{g_2} + 1 + \mu }}$, $T_0 = 2^{R_0} - 1$, $T_1 = 2^{2 R_0} - 1$, ${\theta _0} = \frac{{T_0\left( {1 + \mu } \right)}}{{\bar \gamma }}$ and ${\theta _1} = \frac{{T_1\left( {1 - \rho  + \mu } \right)}}{{\left( {1 - \rho } \right)\bar \gamma }}$. Following the procedure given in \cite{Jain'15}, we derive the primary OP as in (21), given as the top of the next page.
\begin{figure*}[!t]
\begin{align}\label{OP-PN-E}
O{P_p} = \left\{ \begin{array}{l}
\frac{{\gamma \left( {{m_0},{\theta _0}/{\Omega _0}} \right)\gamma \left( {{m_1},{\theta _1}/{\Omega _1}} \right)}}{{\Gamma \left( {{m_0}} \right)\Gamma \left( {{m_1}} \right)}} + \int\limits_0^{{T_1}} {\int\limits_{{\theta _1}}^\infty  {\frac{{{x^{{m_0} - 1}}{z^{{m_1} - 1}}{e^{\frac{{ - x}}{{{\Psi _0}}} - \frac{z}{{{\Omega _1}}}}}\gamma \left( {{m_2},\frac{{\left( {{T_1} - x} \right)\left( {1 + \mu } \right)}}{{{\Omega _2}\bar \gamma \rho \eta \left( {\alpha  - \left( {{T_1} - x} \right)\left( {1 - \alpha } \right)} \right)z}}} \right)}}{{\Gamma \left( {{m_0}} \right)\Gamma \left( {{m_1}} \right)\Gamma \left( {{m_2}} \right)\Psi _0^{{m_0}}\Omega _1^{{m_1}}}}dzdx} } , \; \text{if} \; {T_1} \le \alpha /\left( {1 - \alpha } \right)\\
\frac{{\gamma \left( {{m_0},{\theta _0}/{\Omega _0}} \right)\gamma \left( {{m_1},{\theta _1}/{\Omega _1}} \right)}}{{\Gamma \left( {{m_0}} \right)\Gamma \left( {{m_1}} \right)}} + \left( {1 - \frac{{\gamma \left( {{m_1},{\theta _1}/{\Omega _1}} \right)}}{{\Gamma \left( {{m_1}} \right)}}} \right)\frac{{\gamma \left( {{m_0},\frac{{{T_1} - \alpha /\left( {1 - \alpha } \right)}}{{{\Psi _0}}}} \right)}}{{\Gamma \left( {{m_0}} \right)}} + \int\limits_{{T_1} - \frac{\alpha }{{1 - \alpha }}}^{{T_1}} {\int\limits_{{\theta _1}}^\infty  {\frac{{{x^{{m_0} - 1}}{z^{{m_1} - 1}}{e^{\frac{{ - x}}{{{\Psi _0}}} - \frac{z}{{{\Omega _1}}}}}}}{{\Gamma \left( {{m_0}} \right)\Gamma \left( {{m_1}} \right)\Gamma \left( {{m_2}} \right)\Psi _0^{{m_0}}\Omega _1^{{m_1}}}}} } \\
 \cdot \gamma \left( {{m_2},\frac{{\left( {{T_1} - x} \right)\left( {1 + \mu } \right)}}{{{\Omega _2}\bar \gamma \rho \eta \left( {\alpha  - \left( {{T_1} - x} \right)\left( {1 - \alpha } \right)} \right)z}}} \right), \; \text{if} \; {T_1} > \alpha /\left( {1 - \alpha } \right)
\end{array} \right.
\end{align}
\vspace*{2pt}
\hrulefill
\end{figure*}
In (21), ${\Psi_0} = \frac{{\bar \gamma {\Omega _0}}}{{1 + \mu }}$. Although (\ref{OP-PN-E}) contains double integrals of the lower incomplete Gamma function, it can be readily evaluated by using standard mathematical programs such as Matlab and Mathematica.
\setcounter{equation}{25}
\begin{figure*}[!t]
\begin{align}\label{OP-SN-E}
O{P_s} = \left\{ \begin{array}{l}
\frac{{\gamma \left( {{m_1},{\theta _1}/{\Omega _1}} \right)}}{{\Gamma \left( {{m_1}} \right)}} + \frac{{1 - \gamma \left( {{m_4},{\theta _0}/{\Omega _4}} \right)/\Gamma \left( {{m_4}} \right)}}{{\Gamma \left( {{m_1}} \right)\Gamma \left( {{m_3}} \right)\Omega _1^{{m_1}}}}\int\limits_{{\theta _1}}^\infty  {{x^{{m_1} - 1}}{e^{ - x/{\Omega _1}}}\gamma \left( {{m_3},\frac{{{T_1}\left( {1 + \mu } \right)}}{{{\Omega _3}\bar \gamma \rho \eta \left( {1 - \alpha } \right)x}}} \right)dx}  + \frac{{\gamma \left( {{m_4},{\theta _0}/{\Omega _4}} \right)}}{{\Gamma \left( {{m_4}} \right)}}\\
 \cdot \left( {1 - \frac{{\gamma \left( {{m_1},{\theta _1}/{\Omega _1}} \right)}}{{\Gamma \left( {{m_1}} \right)}}} \right), \; \text{if} \; {T_1} \ge \left( {1 - \alpha } \right)/\alpha \\
\frac{{\gamma \left( {{m_1},{\theta _1}/{\Omega _1}} \right)}}{{\Gamma \left( {{m_1}} \right)}} + \frac{{1 - \gamma \left( {{m_4},{\theta _0}/{\Omega _4}} \right)/\Gamma \left( {{m_4}} \right)}}{{\Gamma \left( {{m_1}} \right)\Gamma \left( {{m_3}} \right)\Omega _1^{{m_1}}}}\int\limits_{{\theta _1}}^\infty  {{x^{{m_1} - 1}}{e^{ - x/{\Omega _1}}}\gamma \left( {{m_3},\frac{{{T_1}\left( {1 + \mu } \right)}}{{{\Omega _3}\bar \gamma \rho \eta \left( {1 - \alpha } \right)x}}} \right)dx} \\
 + \frac{{\gamma \left( {{m_4},{\theta _0}/{\Omega _4}} \right)}}{{\Gamma \left( {{m_1}} \right)\Gamma \left( {{m_3}} \right)\Gamma \left( {{m_4}} \right)\Omega _1^{{m_1}}}}\int\limits_{{\theta _1}}^\infty  {{x^{{m_1} - 1}}{e^{ - x/{\Omega _1}}}\gamma \left( {{m_3},\frac{{{T_1}\left( {1 + \mu } \right)}}{{{\Omega _3}\bar \gamma \rho \eta \left( {1 - \alpha  - \alpha T} \right)x}}} \right)dx} , \; \text{if} \; {T_1} < \left( {1 - \alpha } \right)/\alpha
\end{array} \right.
\end{align}
\vspace*{2pt}
\hrulefill
\end{figure*}

\subsection{Outage Probability of the Secondary System}
The OP of the secondary system can formulated as follows
\setcounter{equation}{21}
\begin{align}\label{OP-SN-1}
O{P_{s}} & = \underbrace {\Pr \left[ {C_{ST}^1 < {R_0}} \right]}_{{J_1}} \nonumber \\
& \;\;\; + \underbrace {\Pr \left[ {C_{ST}^1 \ge {R_0},C_{SR}^1 < {R_0},C_{SR}^{\text{without IC}} < {R_0}} \right]}_{{J_2}} \nonumber \\
& \;\;\; + \underbrace {\Pr \left[ {C_{ST}^1 \ge {R_0},C_{SR}^1 \ge {R_0},C_{SR}^{\text{with IC}} < {R_0}} \right]}_{{J_3}},
\end{align}
where $J_1$ accounts for the probability that ST fails to decode the primary signal, $J_2$ is the probability that ST successfully decodes $x_p$ and SR fails to decode both $x_p$ in the first phase and $x_s$ in the second phase, and $J_3$ denotes the probability that ST and SR successfully decode $x_p$ in the first phase and SR fails to decode $x_s$ in the second phase. Mathematically, $J_1$, $J_2$, and $J_3$ are rewritten as
\begin{align}\label{J1}
{J_1} &= \Pr \left[ {{g_1} < {\theta _1}} \right] = \frac{{\gamma \left( {{m_1},{\theta _1}/{\Omega _1}} \right)}}{{\Gamma \left( {{m_1}} \right)}}, \\
{J_2} &= \Pr \left[ {{g_1} \ge {\theta _1},{g_4} < {\theta_2},\frac{{\left( {1 - \alpha } \right)\bar \gamma \rho \eta {g_1} {g_3}}}{{\alpha \bar \gamma \rho \eta {g_1} {g_3} + 1 + \mu }} < T_1} \right], \\
{J_3} &= \Pr \left[ {{g_4} \ge {\theta_2}} \right]\Pr \left[ {{g_1} \ge {\theta _1},{g_3} < \frac{{T_1 \left( {1 + \mu } \right)}}{{\left( {1 - \alpha } \right)\bar \gamma \rho \eta {g_1}}}} \right].
\end{align}
Following the procedure presented in \cite{Jain'15}, we derive the secondary OP as in (26), shown at the top of this page, where ${\theta _2} = {T_1}\left( {1 + \mu } \right)/\bar \gamma$. It is worthy noting that (\ref{OP-SN-E}) can be readily evaluated by using standard mathematical programs such as Matlab and Mathematica.

\section{High SNR Approximations of Outage Probabilities}
Although the exact expressions of the primary and the secondary OPs given in (\ref{OP-PN-E}) and (\ref{OP-SN-E}) are very useful for quickly evaluating the performance of the primary and the secondary systems, they are, in general, too complex to gain novel insights. This fact motivates us to further investigate the OPs in the high SNR region, from which we can obtain simple and tight approximations of the OPs. In addition, from the approximations we are able to carry out the DOs and the CGs of the two systems.

The DO of a system is defined as $D{O} =  - \mathop {\lim }\limits_{\bar \gamma  \to \infty } \frac{{\log \left( {O{P}} \right)}}{{\log \left( {\bar \gamma } \right)}}$. In addition, in the high SNR regime, we can express a system OP as $O{P} \approx {\left( {\bar \gamma C{G}} \right)^{ - D{O}}}$ \cite{Tse'05}, from which the CG of a system can be carried out. It is noteworthy that DO and CG are the foremost important factors that characterise the asymptotic behaviours of a communication system. More specifically, they provide the information about how fast a system OP decays in the high SNR regime. Moreover, they might provide insights about the effects of key parameters on the system performance.

\subsection{High SNR Analysis for the Primary System}
When $\bar{\gamma} \rightarrow \infty$, we can approximate $I_2$ given in (20) as follows
\setcounter{equation}{26}
\begin{align}\label{Pr2-Approx}
I_2 & \simeq \Pr \left[ {{g_1} \ge {\theta _1},\gamma _{PR}^1 + \frac{\alpha }{{1 - \alpha }} < T_1} \right] \nonumber \\
& = \Pr \left[ {{g_1} \ge {\theta _1}} \right]\Pr \left[ {\gamma _{PR}^1 < T_1 - \frac{\alpha }{{1 - \alpha }}} \right] \nonumber \\
& = \left\{ \begin{array}{l}
0,\; {\text{if}} \; T_1 \le \alpha /\left( {1 - \alpha } \right), \\
\left( {1 - \frac{{\gamma \left( {{m_1},{\theta _1}/{\Omega_1}} \right)}}{{\Gamma \left( {{m_1}} \right)}}} \right) \frac{{\gamma \left( {{m_0},\left( {T_1 - \frac{\alpha }{{1 - \alpha }}} \right)/{\Psi_0}} \right)}}{{\Gamma \left( {{m_0}} \right)}},\\
{\text{if}} \; T_1 > \alpha /\left( {1 - \alpha } \right).
\end{array} \right.
\end{align}
As a result, we obtain the following approximation of the primary OP
\begin{align}\label{OP-PN-Approx}
O{P_p} \simeq \left\{ \begin{array}{l}
\frac{{\gamma \left( {{m_0},{\theta _0}/{\Omega _0}} \right)\gamma \left( {{m_1},{\theta _1}/{\Omega _1}} \right)}}{{\Gamma \left( {{m_0}} \right)\Gamma \left( {{m_1}} \right)}}, \; \text{if} \; {T_1} \le \left( {1 - \alpha } \right)/\alpha \\
\frac{{\gamma \left( {{m_0},{\theta _0}/{\Omega _0}} \right)\gamma \left( {{m_1},{\theta _1}/{\Omega _1}} \right)}}{{\Gamma \left( {{m_0}} \right)\Gamma \left( {{m_1}} \right)}} + \left( {1 - \frac{{\gamma \left( {{m_1},{\theta _1}/{\Omega _1}} \right)}}{{\Gamma \left( {{m_1}} \right)}}} \right)\\
 \cdot \frac{{\gamma \left( {{m_0},\left( {{T_1} - \frac{\alpha }{{1 - \alpha }}} \right)/{\Psi _0}} \right)}}{{\Gamma \left( {{m_0}} \right)}}, \; \text{if} \; {T_1} > \left( {1 - \alpha } \right)/\alpha
\end{array} \right.
\end{align}
which is shown later that it is a very tight approximation over the whole range of $\bar{\gamma}$. To further simplify (\ref{OP-PN-Approx}), an approximation of $\gamma \left( {a,x} \right)$ when $x$ goes to zero is required. We use the series representation of the lower incomplete Gamma function to obtain \cite{Gradshteyn'07}
\begin{align} \label{Gamma-Approx}
\gamma \left( {a,x} \right) = \sum\limits_{m = 0}^\infty  {\frac{{{{\left( { - 1}
\right)}^m}{x^{a + m}}}}{{m!\left( {a + m} \right)}}}  \xrightarrow{x\rightarrow 0} \frac{{{x^a}}}{a}.
\end{align}
Applying (\ref{Gamma-Approx}) to (\ref{OP-PN-Approx}) gives the following results:
\begin{itemize}
  \item When $T_1 \le \alpha /\left( {1 - \alpha } \right)$, we have
    \begin{align}\label{OP-PN-Approx-2}
    OP_{p} &\simeq \frac{{{T_0^{m_0}T_1^{m_1}}{{\left( {\left( {1 + \mu } \right)/{\Omega_0}} \right)}^{{m_0}}}}}{{{m_0}{m_1}\Gamma \left( {{m_0}} \right)\Gamma \left( {{m_1}} \right)}}{\left( {\frac{{1 - \rho  + \mu }}{{{\Omega_1}\left( {1 - \rho } \right)}}} \right)^{{m_1}}} \nonumber \\
    & \;\;\; \cdot \frac{1}{{{{\bar \gamma }^{{m_0} + {m_1}}}}},
    \end{align}
    which shows that the primary DO is $m_0 + m_1$.
  \item When $T_1 > \alpha /\left( {1 - \alpha } \right)$, we have
    \begin{align}\label{OP-PN-Approx-3}
    & OP_{p} \simeq \frac{1}{{{m_0}\Gamma \left( {{m_0}} \right)}}\left( {\frac{{T_1 - \alpha /\left( {1 - \alpha } \right)}}{{{\Omega_0}/\left( {1 + \mu } \right)}}} \right)^{m_0} \frac{1}{{{{\bar \gamma }^{{m_0}}}}} \nonumber \\
    & + \frac{{{T_0^{m_0}T_1^{m_1}}{{\left( {\left( {1 + \mu } \right)/{\Omega_0}} \right)}^{{m_0}}}}}{{{m_0}{m_1}\Gamma \left( {{m_0}} \right)\Gamma \left( {{m_1}} \right)}}{\left( {\frac{{1 - \rho  + \mu }}{{{\Omega_1}\left( {1 - \rho } \right)}}} \right)^{{m_1}}}\frac{1}{{{{\bar \gamma }^{{m_0} + {m_1}}}}} \nonumber \\
    & - \frac{1}{{{m_0}{m_1}\Gamma \left( {{m_0}} \right)\Gamma \left( {{m_1}} \right)}}{\left( {\frac{{T_1 - \alpha /\left( {1 - \alpha } \right)}}{{{\Omega_0}/\left( {1 + \mu } \right)}}} \right)^{{m_0}}} \nonumber \\
    & \cdot {\left( {\frac{{T_1 \left( {1 - \rho  + \mu } \right)}}{{{\Omega_1}\left( {1 - \rho } \right)}}} \right)^{{m_1}}}\frac{1}{{{{\bar \gamma }^{{m_0} + {m_1}}}}} \nonumber \\
    & \simeq \frac{1}{{{m_0}\Gamma \left( {{m_0}} \right)}}\left( {\frac{{T_1 - \alpha /\left( {1 - \alpha } \right)}}{{{\Omega_0}/\left( {1 + \mu } \right)}}} \right)^{m_0} \frac{1}{{{{\bar \gamma }^{{m_0}}}}},
    \end{align}
    which implies that the primary DO is $m_0$.
\end{itemize}
In summary, the primary DO and CG are expressed as follows
\begin{align}\label{DO-PN}
D{O_{p}} = \left\{ \begin{array}{l}
{m_0} + {m_1},\; {\text{if}}\; T_1 \le \alpha /\left( {1 - \alpha } \right), \\
{m_0},\; {\text{if}}\; T_1 > \alpha /\left( {1 - \alpha } \right),
\end{array} \right.
\end{align}
\begin{align}
{CG_p} = \left\{ \begin{array}{l}
{\left[ {\frac{{T_0^{{m_0}}T_1^{{m_1}}{{\left( {\left( {1 + \mu } \right)/{\Omega_0}} \right)}^{{m_0}}}}}{{{m_0}{m_1}\Gamma \left( {{m_0}} \right)\Gamma \left( {{m_1}} \right)}}{{\left( {\frac{{1 - \rho  + \mu }}{{{\Omega_1}\left( {1 - \rho } \right)}}} \right)}^{{m_1}}}} \right]^{\frac{{ - 1}}{{{m_0} + {m_1}}}}}\\
\;\;\;\;\;\;\;\;\;\;\;\;\;\;\;\;\;\;\;\;\;\;\;\;\;\;\;\;\;\;\;\;\;\;\; {\text{if}} \; {T_1} \le \alpha /\left( {1 - \alpha } \right), \\
{\left[ {\frac{1}{{{m_0}\Gamma \left( {{m_0}} \right)}}{{\left( {\frac{{{T_1} - \alpha /\left( {1 - \alpha } \right)}}{{{\Omega_0}/\left( {1 + \mu } \right)}}} \right)}^{{m_0}}}} \right]^{\frac{{ - 1}}{{{m_0}}}}}\\
\;\;\;\;\;\;\;\;\;\;\;\;\;\;\;\;\;\;\;\;\;\;\;\;\;\;\;\;\;\;\;\;\;\;\;  {\text{if}} \; {T_1} > \alpha /\left( {1 - \alpha } \right).
\end{array} \right.
\end{align}
Equation (32) shows that the primary DO solely depends on the fading figures of the PT-PR and the PT-ST channels. In addition, our findings recommend that if the performance of the primary system is the main concern, we should select $\alpha$ so that $\alpha /\left( {1 - \alpha } \right) \ge T_1$.

{\textit{In a special case of Rayleigh fading with $m_j = 1$}, the approximation of the primary OP simplifies to
\begin{align} \label{OP-PN-Ray}
O{P_{p}} \simeq \left\{ \begin{array}{l}
\left( {1 - {e^{ - {\theta _0}/{\beta _0}}}} \right)\left( {1 - {e^{ - {\theta _1}/{\beta _1}}}} \right), \; {\text{if}} \; T_1 \le \alpha /\left( {1 - \alpha } \right),\\
\left( {1 - {e^{ - {\theta _0}/{\beta _0}}}} \right)\left( {1 - {e^{ - {\theta _1}/{\beta _1}}}} \right) + {e^{ - {\theta _1}/{\beta _1}}} \\
\cdot \left( {1 - {e^{\frac{{T_1 - \alpha /\left( {1 - \alpha } \right)}}{{{\Psi_0}}}}}} \right),\; {\text{if}} \; T_1 > \alpha /\left( {1 - \alpha } \right),
\end{array} \right.
\end{align}
\begin{align} \label{DO-PN-Ray}
D{O_{p}} = \left\{ \begin{array}{l}
2,\; {\text{if}}\; T_1 \le \alpha /\left( {1 - \alpha } \right), \\
1,\; {\text{if}}\; T_1 > \alpha /\left( {1 - \alpha } \right),
\end{array} \right.
and
\end{align}
\begin{align} \label{CG-PN-Ray}
{CG_p} = \left\{ \begin{array}{l}
{\left( {\frac{{{T_0}{T_1}\left( {1 + \mu } \right)\left( {1 - \rho  + \mu } \right)/}}{{{\beta _0}{\beta _1}\left( {1 - \rho } \right)}}} \right)^{\frac{{ - 1}}{2}}}, \; {\text{if}} \; {T_1} \le \alpha /\left( {1 - \alpha } \right), \\
{\left( {\frac{{{T_1} - \alpha /\left( {1 - \alpha } \right)}}{{{\beta _0}/\left( {1 + \mu } \right)}}} \right)^{ - 1}}, \; {\text{if}} \; {T_1} > \alpha /\left( {1 - \alpha } \right).
\end{array} \right.
\end{align}
It is worth mentioning that the results given in (\ref{OP-PN-Ray})-(\ref{CG-PN-Ray}) are also presented here for the first time.

\subsection{High SNR Analysis for the Secondary System}
In the high region of $\bar{\gamma}$, we can approximate $J_2$ as follows
\begin{align}\label{J2-Approx}
{J_2} &\simeq \Pr \left[ {{g_4} < {\theta_2}} \right]\Pr \left[ {{g_1} \ge {\theta _1},\frac{{1 - \alpha }}{a} < T_1} \right] \nonumber\\
& = \left\{ \begin{array}{l}
0, \; {\text{if}} \; T_1 \le \left( {1 - \alpha } \right)/\alpha, \\
\left( {1 - \frac{{\gamma \left( {{m_1},{\theta _1}/{\Omega_1}} \right)}}{{\Gamma \left( {{m_1}} \right)}}} \right)\frac{{\gamma \left( {{m_4},{\theta_2}/{\Omega_4}} \right)}}{{\Gamma \left( {{m_4}} \right)}}, \; {\text{if}} \; T_1 > \left( {1 - \alpha } \right)/\alpha.
\end{array} \right.
\end{align}
In addition, an approximation of $J_3$ is given by
\begin{align}
\label{J3-Approx}
{J_3} &= \frac{{1 - \gamma \left( {{m_4},{\theta _2}/{\Omega _4}} \right)/\Gamma \left( {{m_4}} \right)}}{{\Gamma \left( {{m_1}} \right)\Gamma \left( {{m_3}} \right)\Omega _1^{{m_1}}}}\int\limits_{{\theta _1}}^\infty  {{x^{{m_1} - 1}}{e^{ - x/{\Omega _1}}}} \nonumber \\
 & \cdot \gamma \left( {{m_3},\frac{{{\theta _2}}}{{{\Omega _3}\rho \eta \left( {1 - \alpha } \right)x}}} \right)dx \nonumber \\
 & \simeq \frac{{1 - \gamma \left( {{m_4},{\theta _2}/{\Omega _4}} \right)/\Gamma \left( {{m_4}} \right)}}{{\Gamma \left( {{m_1}} \right){m_3}\Gamma \left( {{m_3}} \right)\Omega _1^{{m_1}}}}{\left( {\frac{{{\theta _2}}}{{{\Omega _3}\rho \eta \left( {1 - \alpha } \right)}}} \right)^{{m_3}}} \nonumber \\
 & \cdot \int\limits_{{\theta _1}}^\infty  {{x^{{m_1} - {m_3} - 1}}{e^{ - x/{\Omega _1}}}dx} \nonumber \\
 & = \frac{{1 - \gamma \left( {{m_4},{\theta _2}/{\Omega _4}} \right)/\Gamma \left( {{m_4}} \right)}}{{\Gamma \left( {{m_1}} \right){m_3}\Gamma \left( {{m_3}} \right)\Omega _1^{{m_3}}}}{\left( {\frac{{{\theta _2}}}{{{\Omega _3}\rho \eta \left( {1 - \alpha } \right)}}} \right)^{{m_3}}} \nonumber \\
 & \cdot \Gamma \left( {{m_1} - {m_3},\frac{{{\theta _1}}}{{{\Omega _1}}}} \right),
\end{align}
where the second equation is fulfilled by applying (29) to the first equation and the last equation is satisfied by using [11, 8.350.2]. Finally, taking a summation over (\ref{J1}), (\ref{J2-Approx}), and (\ref{J3-Approx}) gives the following approximation of the secondary OP
\begin{align}\label{OP-SN-Approx}
O{P_s} \simeq \left\{ \begin{array}{l}
\frac{{\gamma \left( {{m_1},{\theta _1}/{\Omega _1}} \right)}}{{\Gamma \left( {{m_1}} \right)}} + \frac{{1 - \gamma \left( {{m_4},{\theta _2}/{\Omega _4}} \right)/\Gamma \left( {{m_4}} \right)}}{{\Gamma \left( {{m_1}} \right){m_3}\Gamma \left( {{m_3}} \right)\Omega _1^{{m_3}}}}\\
{\left( {\frac{{{\theta _2}}}{{{\Omega _3}\rho \eta \left( {1 - \alpha } \right)}}} \right)^{{m_3}}}\Gamma \left( {{m_1} - {m_3},\frac{{{\theta _1}}}{{{\Omega _1}}}} \right), \; \text{if} \; T \le \frac{1-\alpha}{\alpha}\\
\frac{{\gamma \left( {{m_1},{\theta _1}/{\Omega _1}} \right)}}{{\Gamma \left( {{m_1}} \right)}} + \left( {1 - \frac{{\gamma \left( {{m_1},{\theta _1}/{\Omega _1}} \right)}}{{\Gamma \left( {{m_1}} \right)}}} \right)\frac{{\gamma \left( {{m_4},\frac{{{\theta _2}}}{{{\Omega _4}}}} \right)}}{{\Gamma \left( {{m_4}} \right)}}\\
 + \frac{{1 - \gamma \left( {{m_4},{\theta _2}/{\Omega _4}} \right)/\Gamma \left( {{m_4}} \right)}}{{\Gamma \left( {{m_1}} \right){m_3}\Gamma \left( {{m_3}} \right)\Omega _1^{{m_3}}}}{\left( {\frac{{{\theta _2}}}{{{\Omega _3}\rho \eta \left( {1 - \alpha } \right)}}} \right)^{{m_3}}}\\
 \cdot \Gamma \left( {{m_1} - {m_3},\frac{{{\theta _1}}}{{{\Omega _1}}}} \right),\; \text{if} \;T > \frac{1-\alpha}{\alpha}
\end{array} \right.
\end{align}
In order to extract the DO and CG of the secondary system, we need to again utilize the approximation of the lower incomplete Gamma function given in (\ref{Gamma-Approx}). More specifically, applying (\ref{Gamma-Approx}) to (\ref{OP-SN-Approx}) gives:
\begin{itemize}
  \item When $T_1 \le (1 - \alpha)/\alpha$, we have
    \begin{align}\label{OP-SN-Approx-2}
    O{P_s} & \approx \frac{1}{{{{\bar \gamma }^{{m_1}}}{m_1}\Gamma \left( {{m_1}} \right)}}{\left( {\frac{{{T_1}\left( {1 - \rho  + \mu } \right)}}{{\left( {1 - \rho } \right){\Omega _1}}}} \right)^{{m_1}}} \nonumber \\
    & + \frac{{\Gamma \left( {{m_1} - {m_3}} \right)\Omega _1^{ - {m_3}}}}{{{{\bar \gamma }^{{m_3}}}\Gamma \left( {{m_1}} \right){m_3}\Gamma \left( {{m_3}} \right)}}{\left( {\frac{{{T_1}\left( {1 + \mu } \right)}}{{{\Omega _3}\rho \eta \left( {1 - \alpha } \right)}}} \right)^{{m_3}}} \nonumber \\
    &= \frac{1}{{{{\bar \gamma }^{\min \left( {{m_1},{m_3}} \right)}}}}{A_s},
    \end{align}
    where $A_{s}$ is given by
    \begin{align}\label{A-SN}
    {A_s} = \left\{ \begin{array}{l}
    {A_{s1}} = \frac{1}{{{m_1}\Gamma \left( {{m_1}} \right)}}{\left( {\frac{{{T_1}\left( {1 - \rho  + \mu } \right)}}{{\left( {1 - \rho } \right){\Omega _1}}}} \right)^{{m_1}}}, \; \text{if} \; {m_1} < {m_3}\\
    {A_{s2}} = \frac{{\Gamma \left( {{m_1} - {m_3}} \right)\Omega _1^{ - {m_3}}}}{{\Gamma \left( {{m_1}} \right){m_3}\Gamma \left( {{m_3}} \right)}}{\left( {\frac{{{T_1}\left( {1 + \mu } \right)}}{{{\Omega _3}\rho \eta \left( {1 - \alpha } \right)}}} \right)^{{m_3}}}\\
    \;\;\;\;\;\;\;\;\;\;\;\;\;\;\;\;\;\;\;\;\;\;\;\;\;\;\;\;\;\;\;\;\;\;\;\;\;\;\;\;\;\;\;\;\;\;\;\;\; \text{if} \; {m_1} > {m_3}\\
    {A_{s1}} + {A_{s2}},\; \text{if} \; {m_1} = {m_3}
    \end{array} \right.
    \end{align}
    Equation (\ref{OP-SN-Approx-2}) shows that the secondary system can achieve a DO equal to ${\min \left\{ {{m_1},{m_3}} \right\}}$ when $T_1 \le (1 - \alpha)/\alpha$.
  \item When $T_1 > (1 - \alpha)/\alpha$, we have
    \begin{align} \label{OP-SN-Approx-3}
    O{P_s} & \approx \frac{1}{{{{\bar \gamma }^{{m_1}}}{m_1}\Gamma \left( {{m_1}} \right)}}{\left( {\frac{{{T_1}\left( {1 - \rho  + \mu } \right)}}{{\left( {1 - \rho } \right){\Omega _1}}}} \right)^{{m_1}}} \nonumber \\
    &+ \frac{1}{{{{\bar \gamma }^{{m_4}}}{m_4}\Gamma \left( {{m_4}} \right)}}{\left( {\frac{{{T_1}\left( {1 + \mu } \right)}}{{{\Omega _4}}}} \right)^{{m_4}}} \nonumber \\
    &+ \frac{{\Gamma \left( {{m_1} - {m_3}} \right)\Omega _1^{ - {m_3}}}}{{{{\bar \gamma }^{{m_3}}}\Gamma \left( {{m_1}} \right){m_3}\Gamma \left( {{m_3}} \right)}}{\left( {\frac{{{T_1}\left( {1 + \mu } \right)}}{{{\Omega _3}\rho \eta \left( {1 - \alpha } \right)}}} \right)^{{m_3}}} \nonumber \\
    &= \frac{1}{{{{\bar \gamma }^{\min \left( {{m_1},{m_3},{m_4}} \right)}}}}{B_s},
    \end{align}
    where $B_{s}$ is given by
    \begin{align}\label{B-SN}
    {B_s} = \left\{ \begin{array}{l}
    {B_{s1}} = \frac{1}{{{m_1}\Gamma \left( {{m_1}} \right)}}{\left( {\frac{{{T_1}\left( {1 - \rho  + \mu } \right)}}{{\left( {1 - \rho } \right){\Omega _1}}}} \right)^{{m_1}}},\\
    \;\;\;\;\;\;\;\;\;\ \text{if} \; {m_1} < \min \left\{ {{m_3},{m_4}} \right\}\\
    {B_{s2}} = \frac{{\Gamma \left( {{m_1} - {m_3}} \right)\Omega _1^{ - {m_3}}}}{{\Gamma \left( {{m_1}} \right){m_3}\Gamma \left( {{m_3}} \right)}}{\left( {\frac{{{T_1}\left( {1 + \mu } \right)}}{{{\Omega _3}\rho \eta \left( {1 - \alpha } \right)}}} \right)^{{m_3}}},\\
    \;\;\;\;\;\;\;\;\;\ \text{if} \; {m_3} < \min \left\{ {{m_1},{m_4}} \right\}\\
    {B_{s3}} = \frac{1}{{{m_4}\Gamma \left( {{m_4}} \right)}}{\left( {\frac{{{T_1}\left( {1 + \mu } \right)}}{{{\Omega _4}}}} \right)^{{m_4}}},\\
    \;\;\;\;\;\;\;\;\;\ \text{if} \; {m_4} < \min \left\{ {{m_1},{m_3}} \right\}\\
    {B_{s1}} + {B_{s2}}, \; \text{if} \; {m_1} = {m_3} < {m_4}\\
    {B_{s1}} + {B_{s3}}, \; \text{if} \; {m_1} = {m_4} < {m_3}\\
    {B_{s2}} + {B_{s3}}, \; \text{if} \; {m_3} = {m_4} < {m_1}\\
    {B_{s1}} + {B_{s2}} + {B_{s3}}, \; \text{if} \; {m_1} = {m_3} = {m_4}
\end{array} \right.
    \end{align}
    Equation (\ref{OP-SN-Approx-3}) points out that the secondary DO is ${\min \left\{ {{m_1},{m_3},m_4} \right\}}$ when $T_1 > (1 - \alpha)/\alpha$.
\end{itemize}
In short, the secondary DO and CG are
\begin{align}\label{DO-SN}
D{O_{s}} = \left\{ \begin{array}{l}
\min \left( {{m_1},{m_3}} \right), \; {\text{if}} \; T_1 \le \left( {1 - \alpha } \right)/\alpha, \\
\min \left( {{m_1},{m_3},{m_4}} \right), \; {\text{if}} \; T_1 > \left( {1 - \alpha } \right)/\alpha,
\end{array} \right.
\end{align}
\begin{align}
{CG_s} = \left\{ \begin{array}{l}
{\left( {{A_s}} \right)^{\frac{{ - 1}}{{\min \left\{ {{m_1},{m_3}} \right\}}}}}, \; {\text{if}} \; {T_1} \le \left( {1 - \alpha } \right)/\alpha, \\
{\left( {{B_s}} \right)^{\frac{{ - 1}}{{\min \left\{ {{m_1},{m_3},{m_4}} \right\}}}}}, \; {\text{if}} \; {T_1} > \left( {1 - \alpha } \right)/\alpha.
\end{array} \right.
\end{align}
Equation (\ref{DO-SN}) presents that the secondary DO depends on the fading figures of the PT-ST, ST-SR, and the PT-SR channels. In addition, the above results suggest that if the secondary performance is more preferable than that of the primary system, we should select $\alpha$ such that $\left( {1 - \alpha } \right)/\alpha \ge T_1$.

\textit{In a special case of Rayleigh fading}, the approximation of the secondary OP simplifies to
\begin{align}
O{P_{s}} \simeq \left\{ \begin{array}{l}
1 - {e^{ - \frac{\theta_1}{\beta_1}}} + \frac{{{e^{ - {\theta_2}/{\beta_4}}}T_1 \left( {1 + \mu } \right)}}{{\bar \gamma {\beta _1}{\beta _3}\left( {1 - \alpha } \right)\rho \eta }}, \; {\text{if}} \; T_1 \le \frac{1-\alpha}{\alpha}, \\
1 - {e^{ - {\theta _1}/{\beta _1}}}{e^{ - {\theta_2}/{\beta_4}}} + \frac{{{e^{ -{\theta_2}/{\beta _4}}}T_1 \left( {1 + \mu } \right)}}{{\bar \gamma {\beta _1}{\beta _3}\left( {1 - \alpha } \right)\rho \eta }}, \\
\;\;\;\;\;\;\;\;\;\;\;\;\;\;\;\;\;\;\;\;\;\;\;\;\;\;\;\;\;\;\;\;\;\;\;\;\;\;\;\;\;\;\;\; {\text{if}} \; T_1 > \frac{1-\alpha}{\alpha}.
\end{array} \right.
\end{align}
In addition, the secondary DO in this special case is unity and the secondary CG gain is given by
\begin{align}
{CG_s} = \left\{ \begin{array}{l}
{\left[ {\frac{{{T_1}\left( {1 - \rho  + \mu } \right)}}{{{\beta _1}\left( {1 - \rho } \right)}} + \frac{{{T_1}\left( {1 + \mu } \right)}}{{\rho \eta {\beta _1}{\beta _3}\left( {1 - \alpha } \right)}}} \right]^{ - 1}}, \; {\text{if}} \; T_1 \le \frac{1-\alpha}{\alpha}, \\
{\left[ {\frac{{{T_1}\left( {1 - \rho  + \mu } \right)}}{{{\beta _1}\left( {1 - \rho } \right)}} + \frac{{{T_1}\left( {1 + \mu } \right)}}{{\rho \eta {\beta _1}{\beta _3}\left( {1 - \alpha } \right)}} + \frac{{{T_1}\left( {1 + \mu } \right)}}{{{\beta _4}}}} \right]^{ - 1}},\\
\;\;\;\;\;\;\;\;\;\;\;\;\;\;\;\;\;\;\;\;\;\;\;\;\;\;\;\;\;\;\;\;\;\;\;\;\;\;\;\;\;\;\;\;\;\;\;\;\; {\text{if}} \; T_1 > \frac{1-\alpha}{\alpha}.
\end{array} \right.
\end{align}
To the best of authors' knowledge, the results given in (46) and (47) are also presented here for the first time.

\section{Simulation Results}
In this section, representative simulated and analytical results are provided to validate our analysis and to examine the impact of the Nakagami-{\textit{m}} fading figures, power sharing coefficient, and the power splitting coefficient on the OPs of the primary and the secondary systems. The simulation setting follows the network model given in Section II with $R_0 = 1$ bits/s/Hz, $\eta = 1$, and $\mu = 1$. In addition, we consider a practical scenario where the direct PT-PR link undergoes severe fading. This scenario is typical for a win-win cooperation between PUs and SUs. That is, due to severe fading, the primary system itself cannot achieve a satisfactory quality-of-service. In this case, SUs can help PUs to obtain a better performance, and in return, SUs can access the licensed band to transmit their own information. Moreover, we also assume that the quality of PT-SR channel is worse than that of PT-ST, ST-PR, and ST-SR channels. The reason is that SR is likely to be located far away from PT.
\begin{figure}[t]
    \centering
    \includegraphics[width = 8cm]{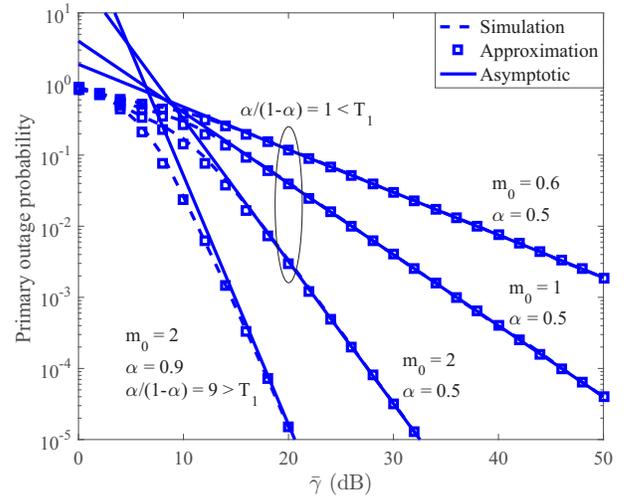}
    \caption{Primary OP versus $\bar{\gamma}$ with $\rho = 0.5$, $m_1 = m_2 = m_3 = 1.5$, $m_4 = 0.6$, $\beta_0 = 1$, $\beta_1 = \beta_2 = \beta_3 = 1.5$, $\beta_4 = 1$, and $T_1 = 3$.}
\end{figure}

In Fig. 3, we simulate the primary OP versus $\bar{\gamma}$. In the figure, the exact and the approximation curves are obtained by using (\ref{OP-PN-E}) and (\ref{OP-PN-Approx}). In addition, the curve representing the primary DO is obtained from (\ref{OP-PN-Approx-2}) and (\ref{OP-PN-Approx-3}). The figure first validates our exact expression of the OP. Secondly, it is shown that the approximation of the OP follows the exact one closely over the whole range of $\bar{\gamma}$ despite the fact that a high SNR approximation is used in our derivation. Thirdly, we observe that the DO curves quickly converge to the exact ones, which validates our findings on the primary DO and CG. For the first three upper curves, we set $\alpha = 0.5$. Since $\alpha/(1 - \alpha) < T_1$, the system DO in this case solely depends on the fading figure of the PT-PR link, $m_0$. Consequently, increasing $m_0$ significantly enhances the primary OP and DO. For the left-most curve, we set $\alpha = 0.9$. Since $\alpha/(1 - \alpha) > T_1$, the primary OP and DO in this case are better than that of the aforementioned case.
\begin{figure}[t]
    \centering
    \includegraphics[width = 8cm]{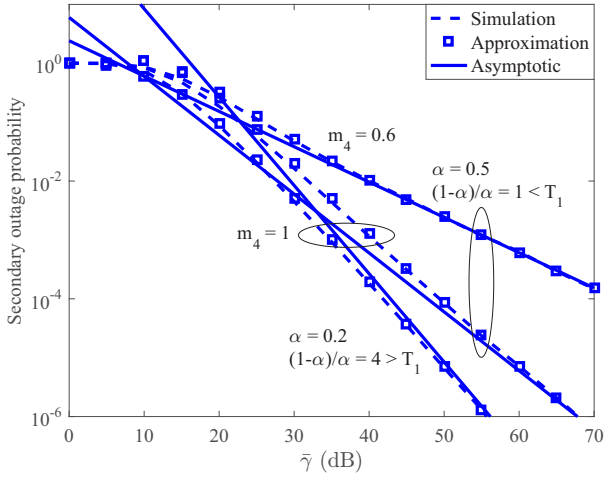}
    \caption{Secondary OP versus $\bar{\gamma}$ with $\rho = 0.5$, $m_0 = 0.6$, $m_1 = m_2 = m_3 = 1.5$, $\beta_0 = 1$, $\beta_1 = \beta_2 = \beta_3 = 1.5$, $\beta_4 = 1$, and $T_1 = 3$.}
\end{figure}
\begin{figure}[t]
 \centering
 \includegraphics[width = 8cm]{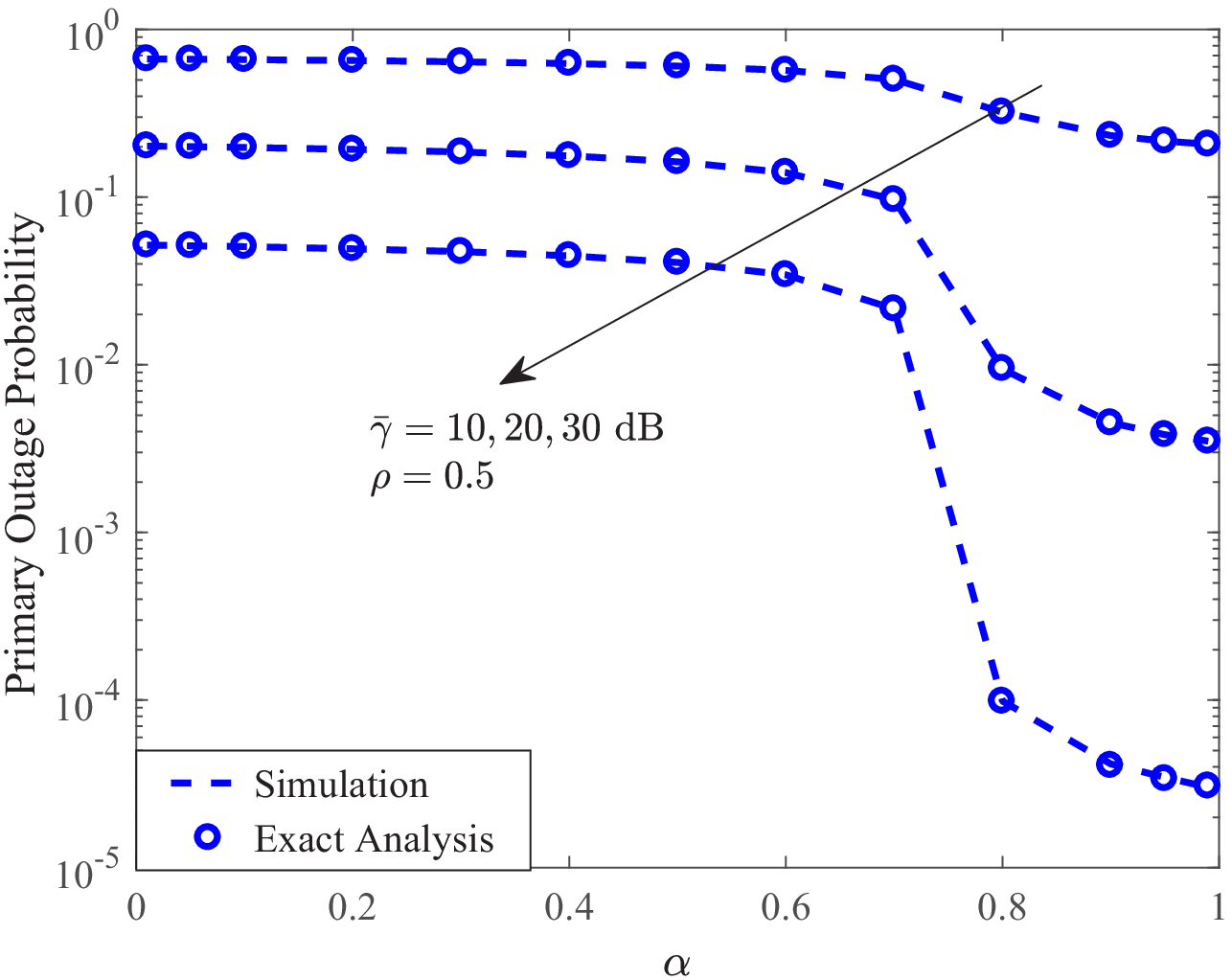}
 \caption{Primary OP versus $\alpha$ with $\rho = 0.5$, $m_0 = 0.6$, $m_1 = m_2 = m_3 = 1.5$, $m_4 = 0.6$, $\beta_0 = 1$, $\beta_1 = \beta_2 = \beta_3 = 1.5$, $\beta_4 = 1$, and $T_1 = 3$.}
\end{figure}
\begin{figure}[t]
 \centering
 \includegraphics[width = 8cm]{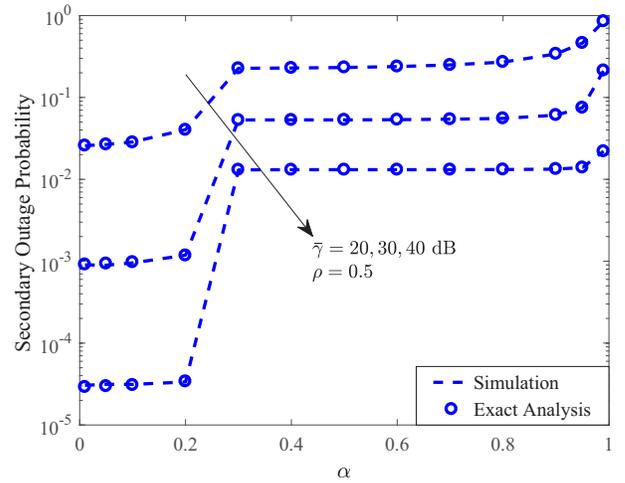}
 \caption{Secondary OP versus $\alpha$ with $m_0 = 0.6$, $m_1 = m_2 = m_3 = 1.5$, $m_4 = 0.6$, $\beta_0 = 1$, $\beta_1 = \beta_2 = \beta_3 = 1.5$, $\beta_4 = 1$, $\rho = 0.5$, and $T_1 = 3$.}
\end{figure}

In Fig. 4, we illustrate the secondary OP versus $\bar{\gamma}$. The exact, approximation, and the DO curves are carried out by (\ref{OP-SN-E}), (\ref{OP-SN-Approx}), (\ref{OP-SN-Approx-2}), and (\ref{OP-SN-Approx-3}), respectively. The figure confirms the validity of our exact expression of the secondary OP. In addition, our approximation of the OP is shown to follows the exact one tightly in the medium-high SNR region. Moreover, the figure also confirms our finding on the secondary DO since the curves representing the DO converge to the others in the high SNR region. For the two upper curves, with $\alpha = 0.5$ and $(1 - \alpha)/\alpha < T_1$, increasing $m_4$ improves the DO. For the two lower curves, reducing $\alpha$ from $0.5$ to $0.2$ also enlarges the DO. It is because when $\alpha = 0.2$, we have $(1 - \alpha)/\alpha > T_1$, and hence, the DO is equal to $\min \left\{ {{m_1},{m_3}} \right\} \ge \min \left\{ {{m_1},{m_3},m_4} \right\}$.

An illustration of the effect of $\alpha$ on the primary OP is given in Fig. 5. It is first shown that the larger $\bar{\gamma}$ the lower the OP, as expected. Secondly, the figure presents that up to a certain value of $\alpha$, i.e., 0.6, the effect of $\alpha$ on the primary system performance is negligible. However, the OP of the primary system sharply decreases when $\alpha$ enlarges from $0.7$ to $0.8$. In addition, the larger $\bar{\gamma}$, the more sharply the improvement. The reason is that at $\alpha = T_1/(1 + T_1) = 0.75$, the primary DO increases from $m_0$ to $m_0 + m_1$, which causes a sudden reduction in the OP.

The effect of $\alpha$ on the secondary OP is illustrated in Fig. 6. Similar to the trend in Fig. 5, the larger $\bar{\gamma}$ the lower the OP. In addition, we observe that increasing $\alpha$ degrades the OP. This result is understandable because the larger $\alpha$ the less power is allocated for the secondary signal, and thus, the received SNR at SR is likely to be not sufficient for a successful decoding. Moreover, it is noted that the secondary OP sharply increases when $\alpha$ enlarges from $0.2$ to $0.3$. The reason is that as $\alpha \geq 1/(1 + T_1) = 0.25$, the system DO reduces from $\min \left\{ {{m_1},{m_3}} \right\}$ to $\min \left\{ {{m_1},{m_3}, m_4} \right\}$, which significantly increases the OP.

We now consider the effect of $\rho$ on the performance of the two systems. The impact of $\rho$ on the primary OP is given in Fig. 7. We see that as $\rho$ approaches zero and one, the OP of the primary system is degraded. The explanation is as follows. When $\rho$ tends to zero, a large fraction of the received power at ST is used for information processing, and thus, ST successfully decodes the primary signal with a large probability. Hence, ST is likely to transmit in the second phase. However, the transmission of ST is mainly supported by its small internal power, which causes a poor primary performance. On the other hand, as $\rho$ approaches one, ST can harvest more power, but may fail to decode the primary signal. Consequently, the primary system relies on the direct PT-PR transmission which takes place over the low quality channel. Therefore, the primary performance is degraded.
\begin{figure}[t]
 \centering
 \includegraphics[width = 8cm]{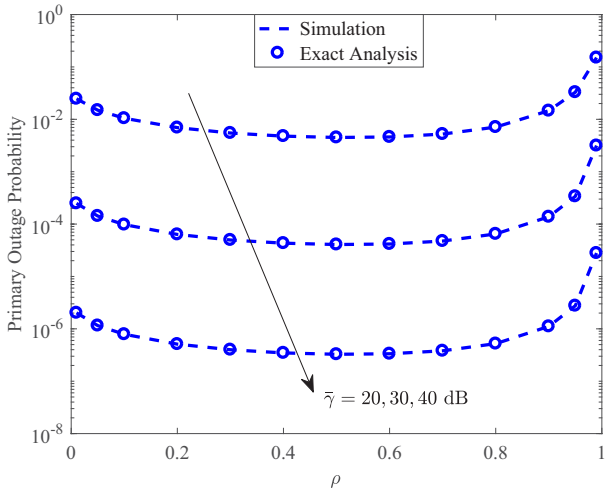}
 \caption{Primary OP versus $\rho$ with $m_0 = 0.6$, $m_1 = m_2 = m_3 = 1.5$, $m_4 = 0.6$, $\beta_0 = 1$, $\beta_1 = \beta_2 = \beta_3 = 1.5$, $\beta_4 = 1$, $\alpha = 0.9$, and $T_1 = 3$.}
\end{figure}
\begin{figure}[t]
 \centering
 \includegraphics[width = 8cm]{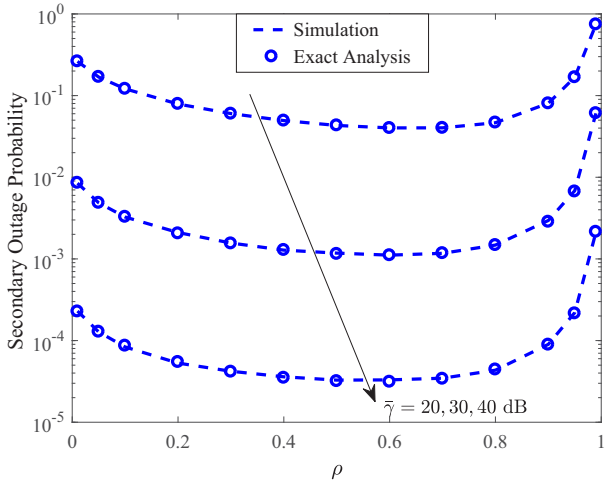}
 \caption{Secondary OP versus $\rho$ with $m_0 = 0.6$, $m_1 = m_2 = m_3 = 1.5$, $m_4 = 0.6$, $\beta_0 = 1$, $\beta_1 = \beta_2 = \beta_3 = 1.5$, $\beta_4 = 1$, $\alpha = 0.2$, and $T_1 = 3$.}
\end{figure}

Fig. 8 shows the effect of $\rho$ on the performance of the secondary system. It is shown that $\rho$ has similar effect on both the primary and the secondary systems. The explanation here is also similar to the one presented in the above paragraph. One important thing we observe from the last two figures is that practical values of $\rho$, at which the primary and the secondary systems can experience good performance simultaneously, lie in the range of $0.5 - 0.7$.

\section{Conclusion}
In this work, we considered an overlay CRN with a RF-based EH ST. We first derived the OPs of the primary and the secondary systems under independent Nakagami-{\textit{m}} fading in integral-based expressions. Thereafter, we obtained very tight closed-form approximations of the OPs, from which we carried out the DOs and CGs of the two systems. We proved that the primary DO is $m_0 + m_1$ if $T_1 \le \alpha / (1 - \alpha)$ or $m_0$ otherwise. In addition, the secondary DO is $\min \left\{ {{m_1},{m_3}} \right\}$ if $T_1 \le (1 - \alpha)/\alpha$ or $\min \left\{ {{m_1},{m_3},m_4} \right\}$ otherwise. Our results also pointed out that the power sharing coefficient plays an important role in deciding the asymptotic performance of the two systems.

%
%
%\newpage
\section*{Author biographies}
\begin{IEEEbiography}[{\includegraphics[width=1in,height=1.25in,clip,keepaspectratio]{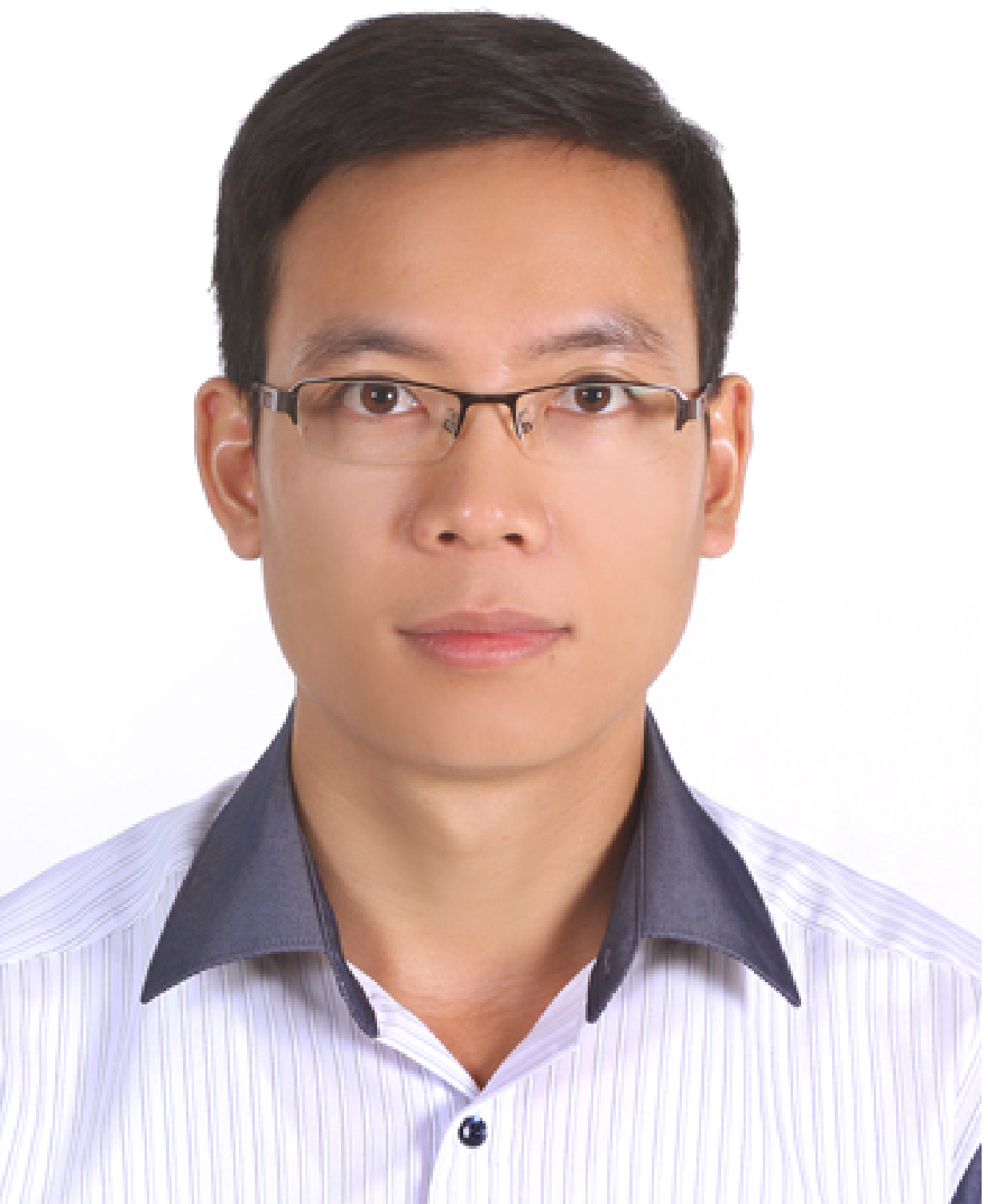}}]
{Binh Van Nguyen} received the B.S. degree from Ho Chi Minh City University of Technology, Ho Chi Minh City, Vietnam, in 2010, and the M.S. and the Ph.D. degrees from Gwangju Institute of Science
and Technology (GIST), Gwangju, South Korea, in 2012 and 2016, respectively, all in wireless communications. He is currently a Research Fellow at GIST. His research interests include cooperative communications, physical layer security, chaotic and anti-jamming communications, and biomedical signal processing and application design.
\end{IEEEbiography}
\begin{IEEEbiography}[{\includegraphics[width=1in,height=1.25in,clip,keepaspectratio]{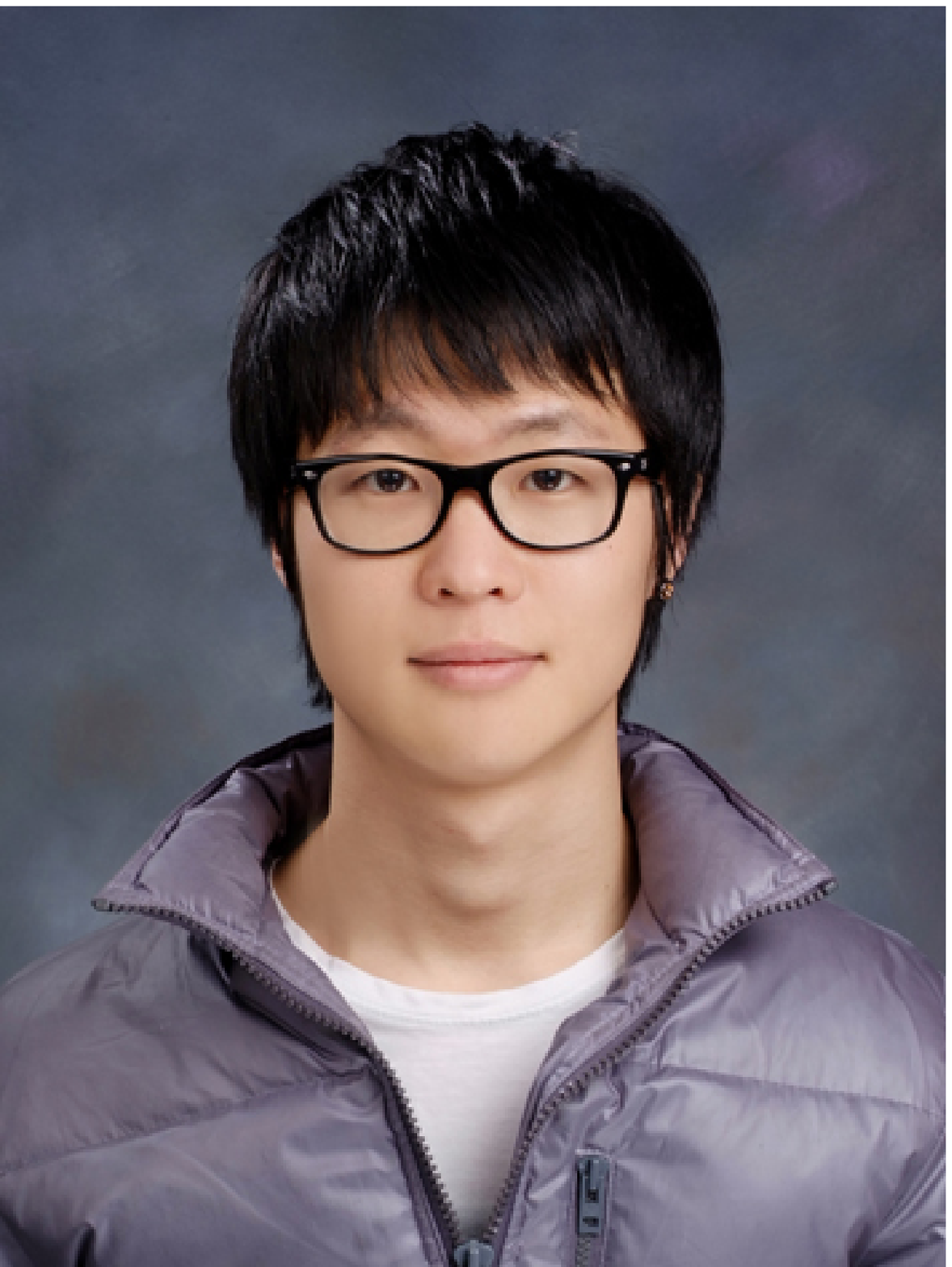}}]
{Hyoyoung Jung} received the B.S. and the M.S. degrees from Inha University and Gwangju Institute of Science and Technology (GIST), South Korea, in 2011 and 2013, respectively. He is currently pursuing the Ph.D. degree at GIST. His research interests include information theory, estimation theory, and time-hopping and anti-jamming communication systems.
\end{IEEEbiography}
\begin{IEEEbiography}[{\includegraphics[width=1in,height=1.25in,clip,keepaspectratio]{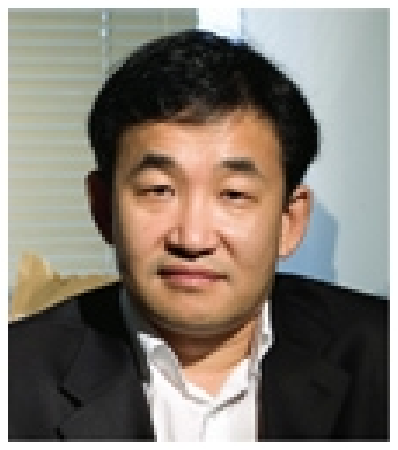}}]
{Dongsoo Har} received the B.S. and M.S. degrees in electronics engineering from Seoul National University, Korea, and the Ph.D. degree in electrical engineering from Polytechnic University, Brooklyn, NY, in 1986, 1988, and 1997, respectively. He was with Vodafone AirTouch, Walnut Creek, CA, from 1997 to 2000, as a Senior Radio Frequency Engineer. From 2003 to 2012, he was with the Department of Information and Communications, Gwangju Institute of Science and Technology, Korea, as an Associate Professor. Since 2013, he has been with the Korea Advanced Institute of Science and Technology, Daejeon, Korea, as a Professor. His current research interests include wireless and wired data network planning, capacity enhancement for multimedia service systems, smart antennas, and system-on-chip designs for multimedia systems. He was a recipient of the IEEE TRANSACTIONS ON VEHICULAR TECHNOLOGY Best Paper Award (Jack Neubauer Memorial Award) in 2000.
\end{IEEEbiography}
\begin{IEEEbiography}[{\includegraphics[width=1in,height=1.25in,clip,keepaspectratio]{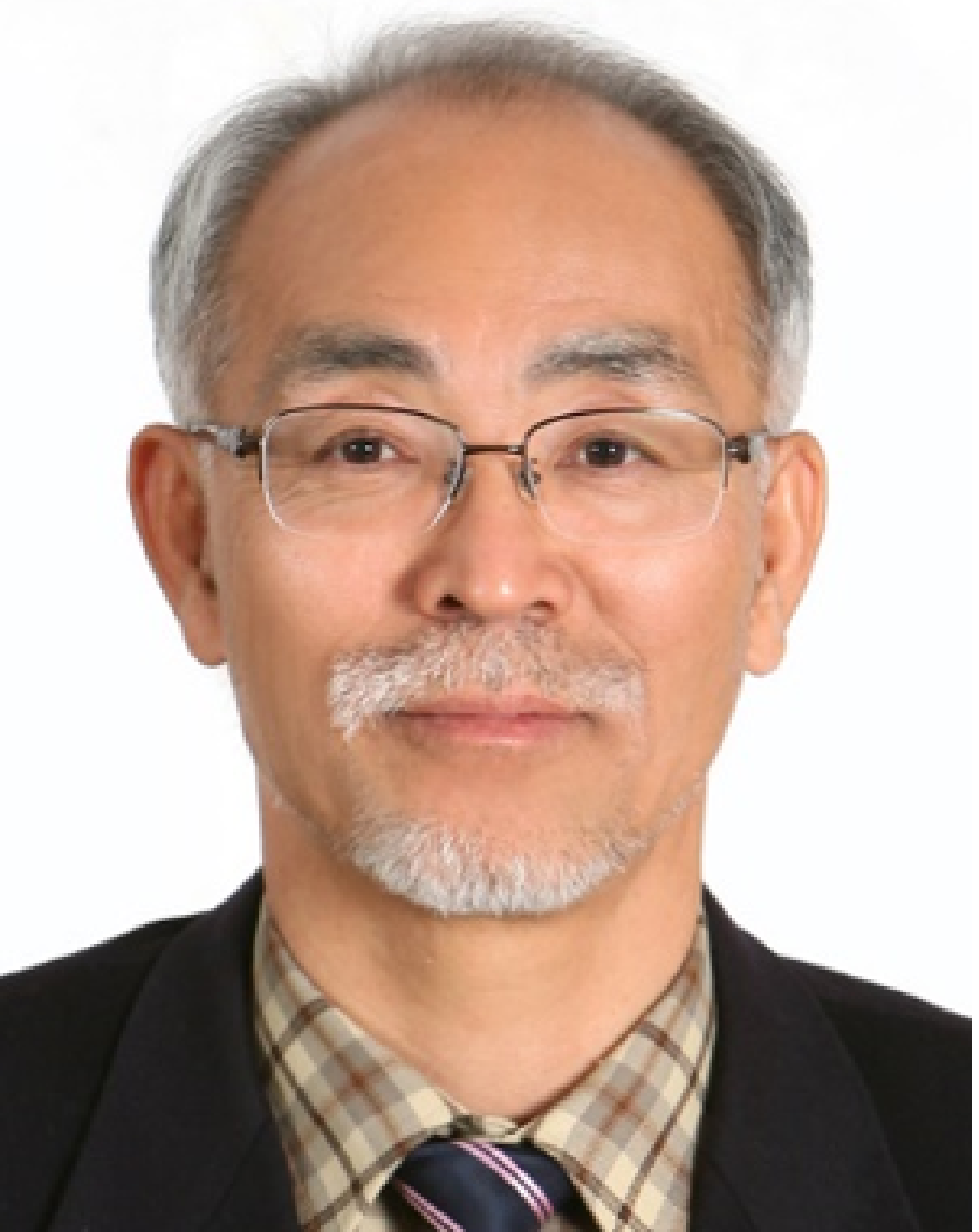}}]
{Kiseon Kim} received the B. Eng and M. Eng. Degrees in electronics engineering from Seoul National Unversity, Seoul, South Korea, in 1978 and 1980, respectively, and the Ph.D. degree in electrical engineering systems from the University of Southern California, Los Angeles, CA, USA, in 1987. From 1988 to 1991, he was with Schlumberger, Houston, TX, USA. From 1991 to 1994, he was with the Superconducting Super Collider Lab, TX, USA. In 1994, he joined Gwangju Institute of Science and Technology, Gwangju, South Korea, where he is currently a Professor. His current research interests include wideband digital communications system design, sensor network design, analysis and implementation both at the physical and at the resource management layer, and biomedical application design. Dr. Kim is the member of the National Academy of Engineering of Korea, the Fellow IET, and the Senior Editor of the IEEE Sensors Journal.
\end{IEEEbiography}

\end{document}